\documentclass[aps,prd,floats,amsmath,amssymb]{revtex4}
\usepackage{hyperref}
\usepackage{mathrsfs}
\usepackage{amsfonts}
\usepackage{epsfig}
\usepackage{dsfont}
\usepackage{dcolumn}
\usepackage{bm}
\usepackage{amsmath}
\usepackage{amsfonts}
\usepackage{amssymb}
\usepackage{graphicx}
\usepackage{epstopdf}
\usepackage{txfonts}
\usepackage[english]{babel}
\makeatletter
\@ifundefined{textcolor}{} {
 \definecolor{BLACK}{gray}{0}
 \definecolor{WHITE}{gray}{1}
 \definecolor{RED}{rgb}{1,0,0}
 \definecolor{GREEN}{rgb}{0,1,0}
 \definecolor{BLUE}{rgb}{0,0,1}
 \definecolor{CYAN}{cmyk}{1,0,0,0}
 \definecolor{MAGENTA}{cmyk}{0,1,0,0}
 \definecolor{YELLOW}{cmyk}{0,0,1,0}     }
\newcommand\mr{\mathscr}

\newcommand\mb{\mathbb}

\def\m{\mu}
\def\n{\nu}

\def\s{\sigma}
\def\o{\omega}

\def\d{\delta}

\def\c{\cdot}
\def\p{\partial}

\def\lag{\langle}
\def\rag{\rangle}

\def\nn{\nonumber}

\begin{document}

\title{Electric and magnetic screenings of gluons in a model
with dimension-2 gluon condensate}
\bigskip
\bigskip
\author{Fukun Xu$^{1}$}
\email{xufukun@mail.ihep.ac.cn}
\author{Mei Huang$^{1,2}$}
\email{huangm@mail.ihep.ac.cn} \affiliation{$^{1}$ Institute of High
Energy Physics, Chinese Academy of Sciences, Beijing, China }
\affiliation{$^{2}$ Theoretical Physics Center for Science
Facilities, Chinese Academy of Sciences, Beijing, China}
\date{\today }
\bigskip

\begin{abstract}
Electric and magnetic screenings of the thermal gluons are studied
by using the background expansion method in a gluodynamic model with
dimension-2 gluon condensate. At low temperature, the electric and
magnetic gluons are degenerate. With the increasing of temperature,
it is found that the electric and magnetic gluons start to split at
certain temperature $T_0$. The electric screening mass changes
rapidly with temperature when $T>T_0$, and the Polyakov loop
expectation value rises sharply around $T_0$ from zero in the vacuum
to a value around $0.8$ at high temperature. This suggests that the
color electric deconfinement phase transition is driven by electric
gluons. It is also observed that the magnetic screening mass keeps
almost the same as its vacuum value, which manifests that the
magnetic gluons remains confined. Both the screening masses and the
Polyakov loop results are qualitatively in agreement with the
Lattice calculations.

\end{abstract}
\maketitle

\section{Introduction}\label{sec:intro}

QCD vacuum is characterized by spontaneous chiral symmetry breaking
and color confinement. It is expected that chiral symmetry can be
restored and color degrees of freedom can be freed at high
temperature and/or density.

The spontaneous chiral symmetry breaking is well understood by the
dimension-3 quark condensate $\lag\bar{q}q\rag$ \cite{NJL} in the
vacuum, which is the order parameter in the chiral limit when the
current quark mass is zero $m=0$, and the chiral restoration is
characterized by the vanishing of quark condensate.

The mechanism of confinement still remains as a challenge. The
confinement is normally taken as the color singlet nature of the
spectrum. However, the color singlet spectrum nature is not unique
for QCD, but also holds for gauge-Higgs theories in which the gauge
group is spontaneously broken. From the specific feature of QCD
dynamics, the Regge trajectories of hadrons indicate the
string-picture of hadrons, and the confinement can be described by
the string picture of hadrons or the linear potential between two
quarks at large distances, {\it i.e.} $V_{\bar{Q}Q}(R)=\sigma R$
with $\sigma$ the string tension. There have been great efforts in
understanding the emergence of string-like object, {\it e.g.} the
Abrikosov flux tubes \cite{string}, the dual superconductor scenario
induced by monopole condensation \cite{dualSC}, and the center
vortices \cite{vortices}. In the limit of infinite heavy current
quark mass, the flux tube never breaks, and it corresponds to the
scenario of "permanent confinement". From the symmetry point of
view, when the current quark mass goes to infinity $m\rightarrow
\infty$, QCD becomes pure gauge SU(3) theory, which is center
symmetric in the vacuum. The non-vanishing string tension
corresponds to the area law for the Wilson loop, vanishing Polyakov
lines, perimeter-law for the 't Hooft loops or the area-law falloff
for the vortex free energy \cite{Greensite-review}. The
deconfinement phase transition referring to the "permanent
confinement" is characterized by the breaking of center symmetry,
and the usually used order parameter is the Polyakov loop
expectation value $\langle L \rangle $ \cite{Polyakov:1978vu}.

There have been also great efforts in understanding confinement and
deconfinement from low-energy Gluodynamics. Varies of vacuum
condensates provide important information to understand the
non-perturbative dynamics of QCD. For example, the gauge invariant
dimension-4 gluon condensate $\lag g^2G^2\rag$ has been widely
investigated in both QCD sum rules and lattice calculations
\cite{Shifman:1978by,Boyd:1996bx,Schafer:1996wv}, and the
non-vanishing value of the condensate does not signal the breaking
of any symmetry directly, but rather the non-perturbative dynamics
of strongly interacting gluon fields. In last decade, there have
been growing interests in dimension-2 gluon condensates $\lag g^2
A^2\rag $ in SU$(N_c)$ gauge theory
\cite{Lavelle:1988eg,Lavelle:1992yh,Gubarev:2000eu,Gubarev:2000nz,
Verschelde:2001ia,Chetyrkin:1998yr,Kondo:2001nq,Slavnov:2004rz,Boucaud:2001st,
Dudal:2003by,Dudal:2003vv}, with the local dimension-2 operator
\begin{equation}
A^2(x)=\sum_{a=1}^{N_c^2-1}\sum_{\m=1}^4A_{\m}^a(x)A_{\m}^a(x).
\label{eq:AA-intro}\end{equation} The dimension-2 gluon condensate
breaks the property of gauge invariance, and it has been
investigated in varies of gauges. For example, the dimension-2
operator $A^2$ gets a special meaning in the Landau gauge
\cite{Gubarev:2000nz,Boucaud:2001st}, in which the condensate is at
an extremum and plays as a saddle point on its gauge orbit, and a
BRST-invariant mixed gluon-ghost condensate has been introduced in
\cite{Kondo:2001nq}. Though it is not gauge invariant, the growing
interests in the dimension-2 gluon condensate lies in that it is
related to the production of the dynamical gluon mass, and the
possible connection between the minimal value of the $<A^2>_{min}$
and the topological defects ({\it e.g.} the magnetic monopoles
\cite{Gubarev:2000nz}). Furthermore, the dimension-2 gluon
condensate has a more close relation with confinement, the
dimension-2 gluon condensate yields the UV corrections
$\Lambda^2/Q^2$ in the QCD running coupling constant
$\alpha_s(Q^2)$, which leads to the linear potential $\sigma_s R$ at
short distances with $\sigma_s\simeq g_R^2 <A_{\mu}^2>$.

It is of great interest to investigate the behavior of the
dimension-2 gluon condensate at finite temperature and its role in
the deconfinement phase transition.

At zero temperature case the space-time space is symmetric under the
$O(4)$ rotation, {\it i.e.} all Lorentz components of the gauge
field $A_\m$ contribute equally to the vacuum. In the finite
temperature, it is more appropriate to divide the gauge boson into
time-like (electric) and space-like (magnetic) components
\cite{Chernodub:2008kf,Vercauteren:2010rk}. This can be viewed as
the different components of the overall variable, because the
rotational symmetry is broken down to (approximate) $O(3)$ spatial
symmetry as the time direction deduces to a finite volume with
$\beta=1/T$. In fact, as we will show, the electric and magnetic
components are quite nontrivial and behaves quite differently at
finite temperature.

On the other hand, the color screening effect is one of the main
features of the quark-gluon plasma(QGP) and has been widely
investigated in lattice and effective theories
\cite{Gao:1989br,Heller:1997nqa,Kaczmarek:1999mm,Kraemmer:2003gd,
Bowman:2004jm,Kaczmarek:2004gv,Peshier:2006ah,Nakamura:2003pu,
Maezawa:2010vj,Fischer:2003rp,DSE-screening}. Significant evidence
shows that gluon confinement is not affected by a small (physical)
number of light quarks\cite{Bowman:2004jm,Fischer:2003rp} and the
nonperturbative features of QCD are most probably generated in the
gauge sector. It is therefore reasonable to study the behavior of
screening of gluons at finite temperature. Lattice result shows that
the QCD coupling constant strength near the critical temperature
$T_c$ is still of the order of one \cite{Kaczmarek:2004gv}, and the
perturbation theory cannot be applied in this region. Especially in
the regime right above the critical temperature, the nonperturbative
effects are supposed to be important.

Therefore, in this work we extend the pure gluodynamic model with
dimension-2 gluon condensate in the vacuum \cite{Celenza:1986th},
and estimating the electric as well as magnetic screening masses of
gluons at finite temperature. We also investigate the contribution
of the dimension-2 gluon condensate to the deconfinement phase
transition

This paper is structured as follows. In Sec.\ref{sec:D2GC} we
introduce the pure gluodynamic model with dimension-2 gluon
condensate in the vacuum, which was developed by Celenza and Shakin
\cite{Celenza:1986th}. Then in Sec.\ref{sec:D2GC-FiniteT}, we extend
the gluodynamic model to finite temperature and define the electric
and magnetic screening masses from the gluon self-energy tensor. We
give the numerical results of the electric and magnetic screening
masses as well as the Polyakov loop expectation value in
Sec.\ref{sec:res&disc} and give the summary in Sec.\ref{sec:concl}.

\section{The Gluodynamic model with Dimension-2 Gluon Condensate}
\label{sec:D2GC}

In this section, we follow Ref.\cite{Celenza:1986th} to introduce
the Celenza-Shakin model which gives the effective action for pure
gluon system with dimension-2 gluon condensate. As an overall
notation the paper is in the framework of Euclidean space.

The pure gluon part of QCD Lagrangian is described by
\begin{equation}
\mr{L}_{G} = -\frac{1}{4}G_{\mu\nu}^aG^a_{\mu\nu}, \label{eq:lg}
\end{equation}
with
\begin{equation}
G_{\m\n}^a = \p_\m{A_\n^a}-\p_\n{A_\m^a}+gf^{abc}{A_\m^b}{A_\n^c}.
\label{eq:gmunu}
\end{equation}

Motivated by the Nambu--Jona-Lasinio model with quark-antiquark
condensate in the vacuum, which is similar to the BCS pairing
condensation in the superconductor, Celenza-Shakin proposed the
"pairing" of two gluons condenses in the vacuum in
Ref.\cite{Celenza:1986th}. The gluon field can be decomposed into a
condensate field $\mb{A}_\m^a$ and a fluctuating field
$\mathscr{A}_{\mu}^a$ \cite{Celenza:1986th,Li:2004te} as,
\begin{equation}
 A_\mu^a(x):=\mathbb{A}_{\mu}^a+\mathscr{A}_{\mu}^a (x),
 \label{eq:0T-A-expansion-x}
\end{equation}
where $\mb{A}_\m^a$ is macroscopically occupied and independent of
$x$, which carries zero vacuum expectation value, i.e.
$\lag\text{vac}|{\mb{A}_\m^a}|\text{vac}\rag=0$. The Fourier
transformation of Eq. \ref{eq:0T-A-expansion-x} has the form of
\begin{equation}
 A_\mu^a(k):=\mathbb{A}_{\mu}^a (k=0)+\mathscr{A}_{\mu}^a (k)
 \equiv \mathbb{A}_\mu^a+ \mathscr{A}_\mu^a (k),
\label{eq:0T-A-exp-k}
\end{equation}
where the background $\mb{A}_\m^a$ carries only zero momentum mode,
and for simplicity we assume it to be a constant.

By using the expansion Eq.(\ref{eq:0T-A-exp-k}), the gluon part of
the QCD Lagrangian becomes
\begin{eqnarray}
\mr{L}_{G} =&-&\frac{1}{4}[\mr{G_{\m\n}^a}\mr{G^a_{\m\n}}
    +2gf^{abc}{\mr{G}_{\m\n}^a}({\mb{A}_\m^b}{\mr{A}_\n^c}
    +{\mr{A}_\m^b}{\mb{A}_\n^c}+{\mb{A}_\m^b}{\mb{A}_\n^c})
    \nonumber \\
 &+ & g^2 f^{eab}f^{ecd}({\mb{A}_{\m}^a}{\mr{A}_{\n}^b} + {\mr{A}_{\m}^a}{\mb{A}_{\n}^b})
        ({\mb{A}_\m^c}{\mr{A}_\n^d}+{\mr{A}_\m^c}{\mb{A}_\n^d})
        \nonumber \\
    & + & 2g^2f^{eab}f^{ecd}{\mb{A}_\m^a}{\mb{A}_\n^b}({\mb{A}_\m^c}{\mr{A}_\n^d}
    +{\mr{A}_\m^c}{\mb{A}_\n^d})
 + g^2
 f^{eab}f^{ecd}{\mb{A}_\m^a}{\mb{A}_\n^b}{\mb{A}_\m^c}{\mb{A}_\n^d}].
\label{eq:G&GG-0T-exp}\end{eqnarray}

As a further assumption one can treat $\mb{A}_\m^a$ as a classical
variable:
\begin{equation}
\mb{A}_\m^a:=\phi_0\hat{\eta}_\m^a,
\label{eq:0T-A-exp-eta}\end{equation} where $\phi_0$ is constant and
$\hat{\eta}_\m^a$ is a vacuum vector. The vector $\hat{\eta}_\m^a$
has the following properties:
\begin{eqnarray}
\hat{\eta}\equiv\frac{\eta}{|\,\eta|\,},~~\eta
_\m^a\equiv(\eta_4^a,~\vec{\eta}^a), ~~(\hat{\eta}_\m^a)^2 =
1,~~\eta^2=\eta_\m^a\eta_\m^a = 32. \nn\end{eqnarray} The averaging
procedure for an operator $O[\hat{\eta}]$ may be written as
\begin{equation}
 \lag O[\hat{\eta}]\rag_{\hat{\eta}}
    =\frac{\int \prod_{a'} \mb{d} \hat{\eta}_{a'} \d(\hat{\eta}\c\hat{\eta}-1) O[\hat{\eta}]}
          {\int \prod_{a'} \mb{d} \hat{\eta}_{a'} \d(\hat{\eta}\c\hat{\eta}-1)}
\nonumber\end{equation}
Now that the field $\eta_\m^a$ plays as the vacuum degree of freedom, then one can consider
the expectation value this averaging as the vacuum expectation i.e.
\begin{eqnarray}
  \lag\text{vac}|O[A_\m^a]|\text{vac}\rag
        \equiv \lag\text{vac}|O[\mb{A}_\m^a]|\text{vac}\rag
        \equiv \lag O[\hat{\eta}_\m^a] \rag_{\hat{\eta}}
\nn . \end{eqnarray} After taking the expecting value in terms of
$\eta_\m^a$, one gets
\begin{equation}
 \lag{\mb{A}_\m^a}{\mb{A}_\n^b}\rag_{\hat{\eta}} = \frac{\d^{ab}}{8}\frac{\d_{\m\n}}{4}\phi_0^2,
    \qquad \lag{\mb{A}_\m^a}{\mb{A}_\m^a}\rag_{\hat{\eta}}=\phi_0^2.
\label{eq:0T-AA}\end{equation} Actually it is the nonzero
expectation value of the double combination $\mb{A}^2$ plays as an
order parameter representing the existence of condensate but not the
gauge field $A_\m^a$ as one spontaneously has the constraint of
\begin{eqnarray}
  \lag O[(\mb{A}_\m^a)^{odd}] \rag_{\hat{\eta}} = 0.
\end{eqnarray}

Then the Lagrangian after this background expansion becomes
\begin{equation}
\lag\mr{L}\rag_{\hat{\eta}} = -\frac{1}{4}\lag GG\rag_{\hat{\eta}}
=-\frac{1}{4}\left[\mr{GG} + 2m_g^2\mr{A}^2 + 4b\phi_0^4\right],
\label{eq:0T-Lagrangian}
\end{equation}
with \begin{equation} m_g^2=\frac{9}{32}g^2\phi_0^2, ~~~
b=\frac{9}{136}g^2. \end{equation} The gluon gets mass because of
the existence of nonperturbative dimension-2 gluon condensate. We
note that the dimension-four gluon condensate $\lag g^2G^2
\rag_{\hat{\eta}}$ is proportional to dimension-2 gluon condensate
${\lag g^2A^2 \rag}_{\hat{\eta}}^2$.

\section{Electric and magnetic screening at finite temperature}
\label{sec:D2GC-FiniteT}

We now use the Lagrangian in Eq.(\ref{eq:0T-Lagrangian}) as the
effective model of pure gluon system. At finite temperature, the
temporal and spatial direction of the gluon field is in general
different, i.e.
\begin{equation}
\mr{A}:=(\mr{A}_4,\vec{\mr{A}}),
\end{equation}
and the Lagrangian can be written as
\begin{equation}
\lag\mr{L}\rag_{\hat{\eta}} = -{1\over4}{\lag GG \rag_{\hat{\eta}}}
        =-\frac{1}{4}\left[\mr{GG}+2(m_E^2\mr{A}_4^2+m_M^2\vec{\mr{A}}^2)+4b\phi_0^4\right],
\label{eq:FT-Lagrangian}
\end{equation}
In the zero temperature limit, one has $m_E^2=m_M^2\equiv m_g^2$.

By adding the gauge-fixing term in Lagrangian i.e.
$\mr{L}_{\text{fix}}=-\frac{1}{2\xi}(\partial_{\mu}
\mr{A}_{\mu})^2$, one can solve the gluon propagator of the
fluctuating field $\mr{A}_\m^a$ from the equation of
\begin{equation}
 \Big[K^2\d_{\m\n}-(1-1/\xi) K_\m K_\n+m_E^2\d_{44}
 +m_M^2\d_{ij}^{\m\n}\Big]\cdot D_{\n\s}(K)
 =\d_{\m\s}.
\end{equation}
The gluon propagator has the form of
\begin{eqnarray}
  D_{\m\n}(K)=\frac{P_{\m\n}^T}{K^2+m_M^2}
    +\frac{K^2P_{\m\n}^L+\xi\,\Big(m_M^2\d_{44}\!
    +\!K_{\m}K_\n+m_E^2{k_{\m}k_{\n}/{k^2}}\Big)}
    {K^2(K^2+m_E^2)-K_4^2(m_E^2-m_M^2)+\xi(k^2m_M^2+K_4^2m_E^2+m_M^2m_E^2)}.
\label{eq:FT-propg-LG}\end{eqnarray}
In the limit of
$\xi\rightarrow\infty$, i.e. in the unitary gauge, the gluon
propagator takes the form of
\begin{eqnarray}
  D_{\m\n}(K)=\frac{1}{K^2+m_M^2}\left(\d_{ij}-\frac{k_\m k_\n}{k^2}\right)
    +\frac{1}{k^2m_M^2+K_4^2m_E^2+m_E^2m_M^2}
    \left(\d_{44}m_M^2+K_{\m}K_\n+m_E^2\frac{k_{\m}k_{\n}}{k^2}\right).
\label{eq:FT-propg}\end{eqnarray}
In the zero temperature limit
($m_E^2=m_M^2=m_g^2$) it becomes a simple form
\begin{equation}
  D_{\m\n}(K)=\frac{1}{K^2+m_g^2}\left(\d_{\m\n} - \frac{K_\m K_\n}{m_g^2}\right)
\nn\end{equation}

The screening masses are defined as the gluon self-energy tensor
$\Pi_{\m\n}^{ab}(p_4,p)$ at the static limit ($p_4=0$,~
$p\rightarrow0$) \cite{Kapusta89,Bellac96}, and the electric and
magnetic screening masses take the following expressions:
\begin{eqnarray}
  m_E^2\d_{44}\d^{ab}=-\Pi_{44}^{ab}(0,p\rightarrow 0),\qquad
  m_M^2\d_{ij}\d^{ab}=-\Pi_{ij}^{ab}(0,p \rightarrow 0).
\label{eq:gap-eq}\end{eqnarray} Here the gluon self-energy tensor is
with full propagator so that it contains both the perturbative and
the nonperturbative contributions of the interaction of the gauge
field. As it was pointed by some authors that, the above definition
does not yield a gauge invariant definition of the screening masses
in a strict sense.

\begin{figure}
  \includegraphics[width=10cm]{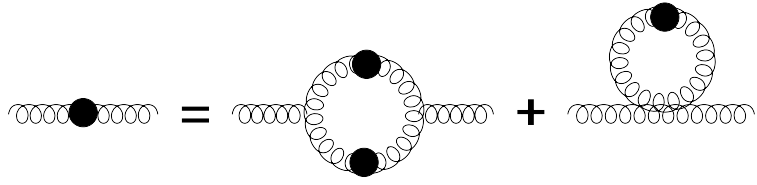}\\
  \caption{Gluon self-energy of Dyson-Schwinger-equation like.}\label{fig:gluon-SE-DSElike}
\end{figure}

On the other hand, we suggest a nonperturbative iterative relation
of gluon mass similar to Dyson-Schwinger method \cite{DSE}, i.e. the
value of screening mass especially at finite temperature is decided
by the gluon self-energy Fig.\ref{fig:gluon-SE-DSElike}

The direct calculation by using propagator Eq.\ref{eq:FT-propg}
gives
\begin{equation}
\Pi_{G,44}^{ab}(P=0)=-g^2N_c\d_{44}\d^{ab}T\sum_n\int{\textbf{d}^3k\over(2\pi)^3}
        \left(2\frac{-\o_n^2+k^2+m_M^2}{(K^2+m_M^2)^2}
          +m_E^2\frac{k^2m_M^2-\o_n^2m_E^2+m_E^2m_M^2}{(k^2m_M^2+\o_n^2m_E^2+m_E^2m_M^2)^2}\right)
\label{eq:E-gluonSE}
\end{equation}
\begin{equation}
\Pi_{G,ij\,}^{ab}(P=0)=
-g^2N_c\d_{i\,j\,}\d^{ab}T\sum_n\int{\textbf{d}^3k\over(2\pi)^3}
\left(2\frac{\o_n^2+k^2/3+m_M^2}{(K^2+m_M^2)^2}
+m_M^2\frac{k^2m_M^2/3+\o_n^2m_E^2+m_E^2m_M^2}{(k^2m_M^2+\o_n^2m_E^2+m_E^2m_M^2)^2}\right)
\label{eq:M-gluonSE}
\end{equation}
Then one immediately gets the results with parameters given.

\section{Results and discussions}
\label{sec:res&disc}

We firstly investigate the thermal behavior of electric and magnetic
screening masses by using the definition Eq.(\ref{eq:gap-eq}) and
the electric and magnetic gluon self-energy in
Eqs.(\ref{eq:E-gluonSE}) and (\ref{eq:M-gluonSE}).

In our model, there are two input parameters, i.e. the
dimension-four gluon condensate $g^2G^2$ or the nonperturbative
coupling constant $g$, and the momentum cutoff parameter $\Lambda$
at zero temperature. For simplicity we assume that the coupling
constant $g$ and cutoff parameter $\Lambda$ remain constants even at
finite temperature. The value of dimension-four gluon condensate at
zero temperature are derived both in QCD sum-rules (lower range of
the interval) \cite{Shifman:1978by,D4GCvalue-SR} and in lattice
(higher range of the interval) \cite{Boyd:1996ex,D4GCvalue-lat}.
Different authors give different results but an acceptable candidate
is $\lag g^2G^2\rag=(0.009\pm0.006)\times4\pi^2 {\text{GeV}}^4$
\cite{Ioffe:2002ee}. We take the value of dimension-four gluon
condensate as $\lag g^2G^2\rag=0.009\times4\pi^2 {\text{GeV}}^4$,
which corresponds to the dimension-2 gluon condensate $\lag
g^2A^2\rag=1.16 {\text{GeV}}^2$ and the gluon mass $m_g=571 {\rm
MeV}$.

For the calculation of the momentum integral, we employ a
soft-cutoff function (for example see \cite{softCutoff}), which
takes the the form of
\begin{equation}
  f(K)=e^{-\Lambda^2K^2}\equiv e^{-\Lambda^2(\o_n^2+k^2)}.
\label{eq:cutoff}\end{equation} In the following numerical
calculation, we choose $\Lambda=0.3 [\text{GeV}^{-1}]$.

\subsection{The electric and magnetic screening masses}
\label{subsec:res&disc-result}

The electric and magnetic screening masses as functions of the
temperature are shown in Fig.\ref{fig:mgT-lam03-xi8}, the solid line
and the dashed-dotted line are for the electric and magnetic part,
respectively.

\begin{minipage}{\textwidth}
\centering
\includegraphics[width=10cm]{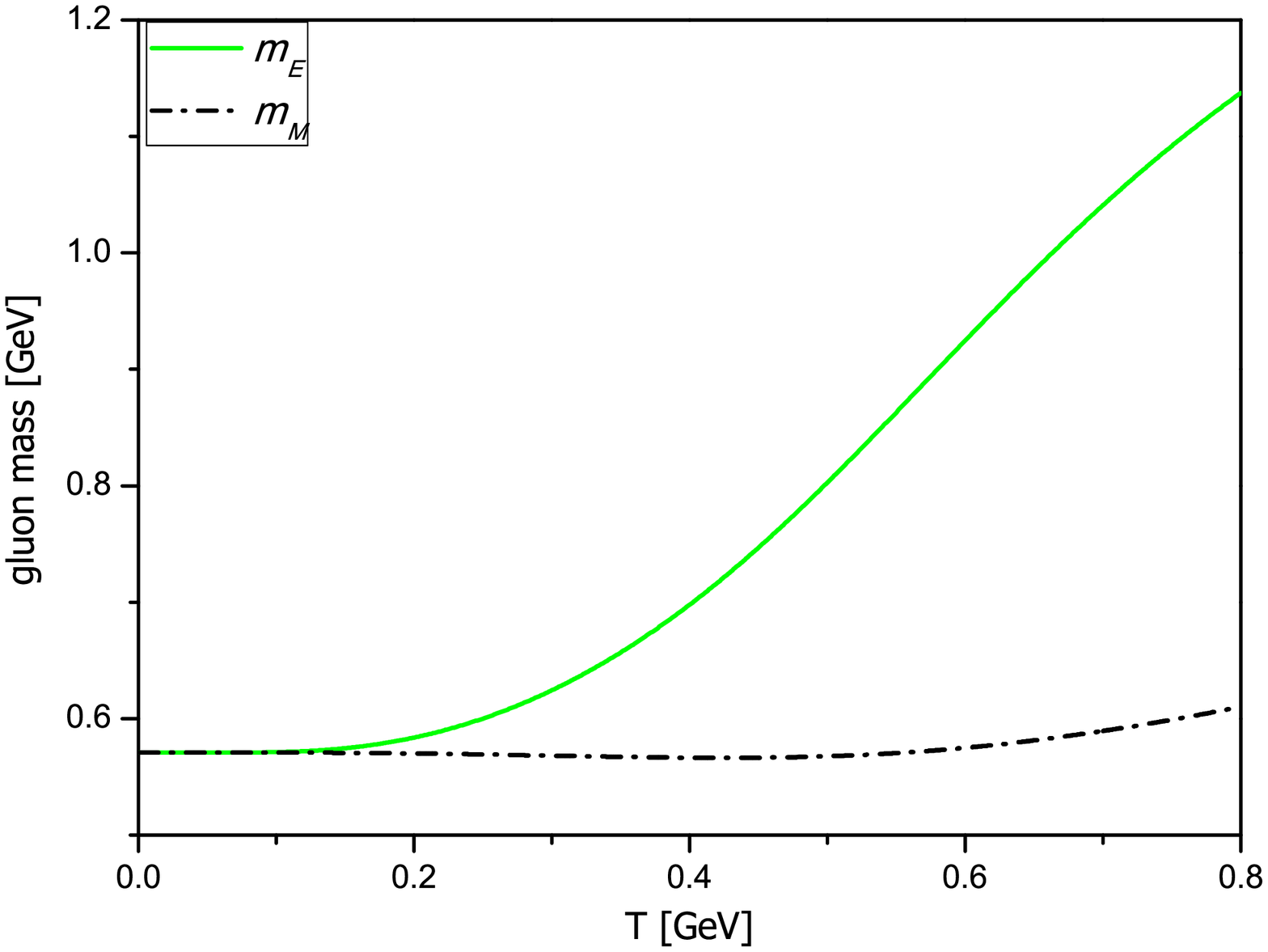}%
\makeatletter\def\@captype{figure}\makeatother \caption{The electric
and magnetic screening masses as functions of the temperature.}
\label{fig:mgT-lam03-xi8}
\end{minipage}
\\[\intextsep]

It is found that both electric and magnetic screening masses are
degenerate and remain unchanged at low temperature, and the electric
and magnetic components start to split at the temperature $T_0=150
{\rm MeV}$. In the temperature region $T>T_0$, the electric
screening mass rise rapidly with the increase of temperature,
however, the magnetic screening mass of the gluons remains almost
the same as its vacuum value.

\begin{figure}[h]
  \includegraphics[width=10cm]{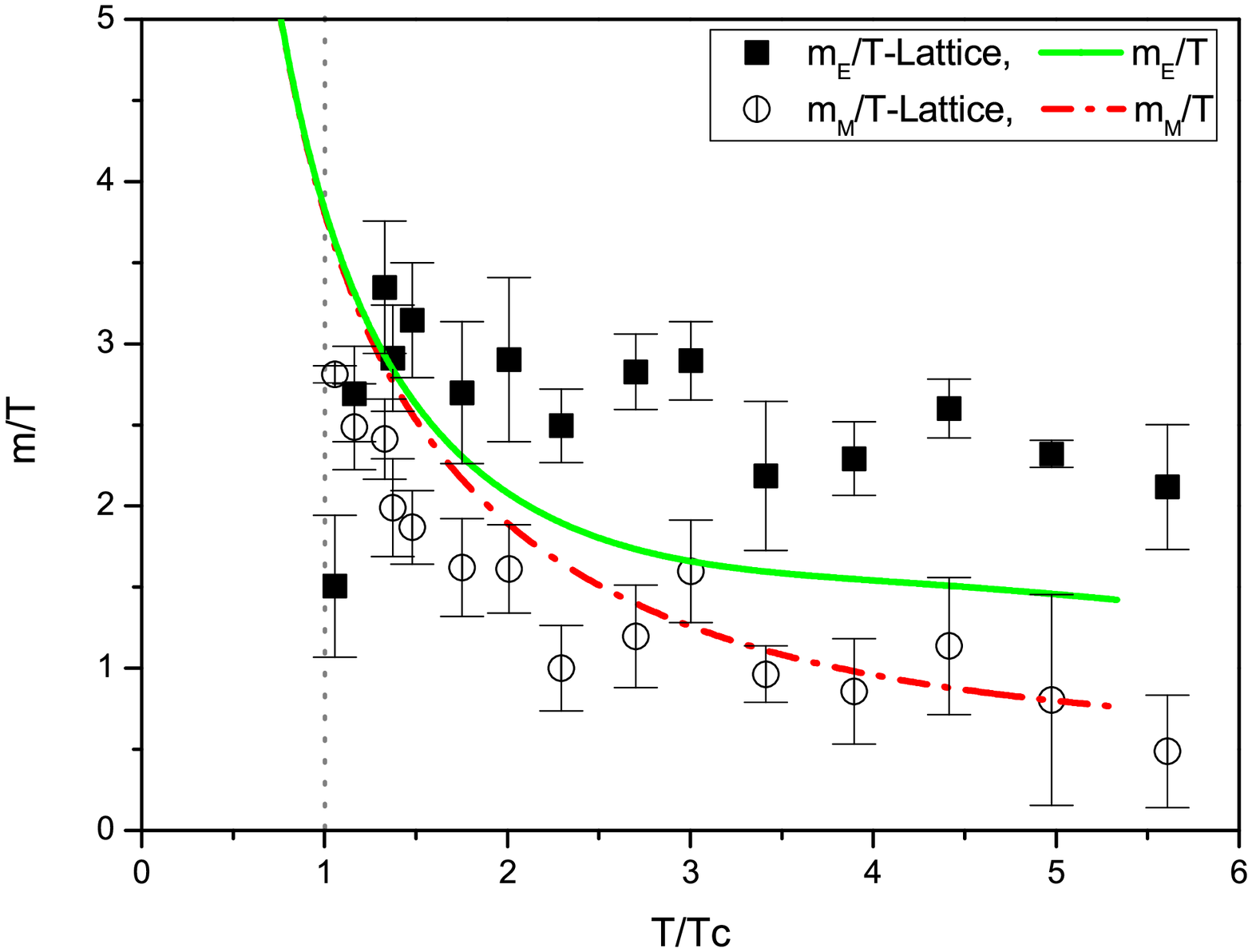}\\
  \caption{The ratios of the screening masses $m_E/T$ and $m_M/T$
as functions of $T/T_c$. The lattice data are taken from
Ref.\cite{Nakamura:2003pu}.}
\label{fig:mgT}
\end{figure}

In order to compare with the lattice data in
Ref.\cite{Nakamura:2003pu}, we divide the screening masses by the
temperature. We also assume the critical temperature $T_c=T_0=150
{\rm MeV}$, where $m_E$ and $m_M$ start to split. (The exact value
of $T_c$ is not important here, and will not affect the qualitative
property of the ratio of the screening mass over the temperature.)
Fig.\ref{fig:mgT} shows the ratios of $m_E/T$ and $ m_M/T$ as
functions of $T/T_c$ and compare with the lattice data in
Ref.\cite{Nakamura:2003pu}. The solid line and the dashed-dotted
line are for the electric and magnetic part, respectively.

It is found that the ratio of the electric screening mass over
temperature $m_E/T$ is around $\sim 1.8$ in the region of
$2<T/T_c<5$, which is qualitatively in agreement with the lattice
result $m_E/T\sim 2.3$. The ratio of the magnetic screening mass
over temperature $m_M/T$ is around  $1$ in the region of
$2<T/T_c<5$, which is almost the same as the lattice result
$m_M/T\sim 1$. It is worthy of mentioning that $m_E/T>m_M/T$ in the
temperature region of $T/T_c<3$ cannot be explained by using the
perturbative scaling $m_E\sim g T$ and $m_M \sim g^2 T$, because of
the coupling constant $g(T)>1$ in this region.

\subsection{Gauge dependence investigation}

\begin{figure}[h]
  \includegraphics[width=10cm]{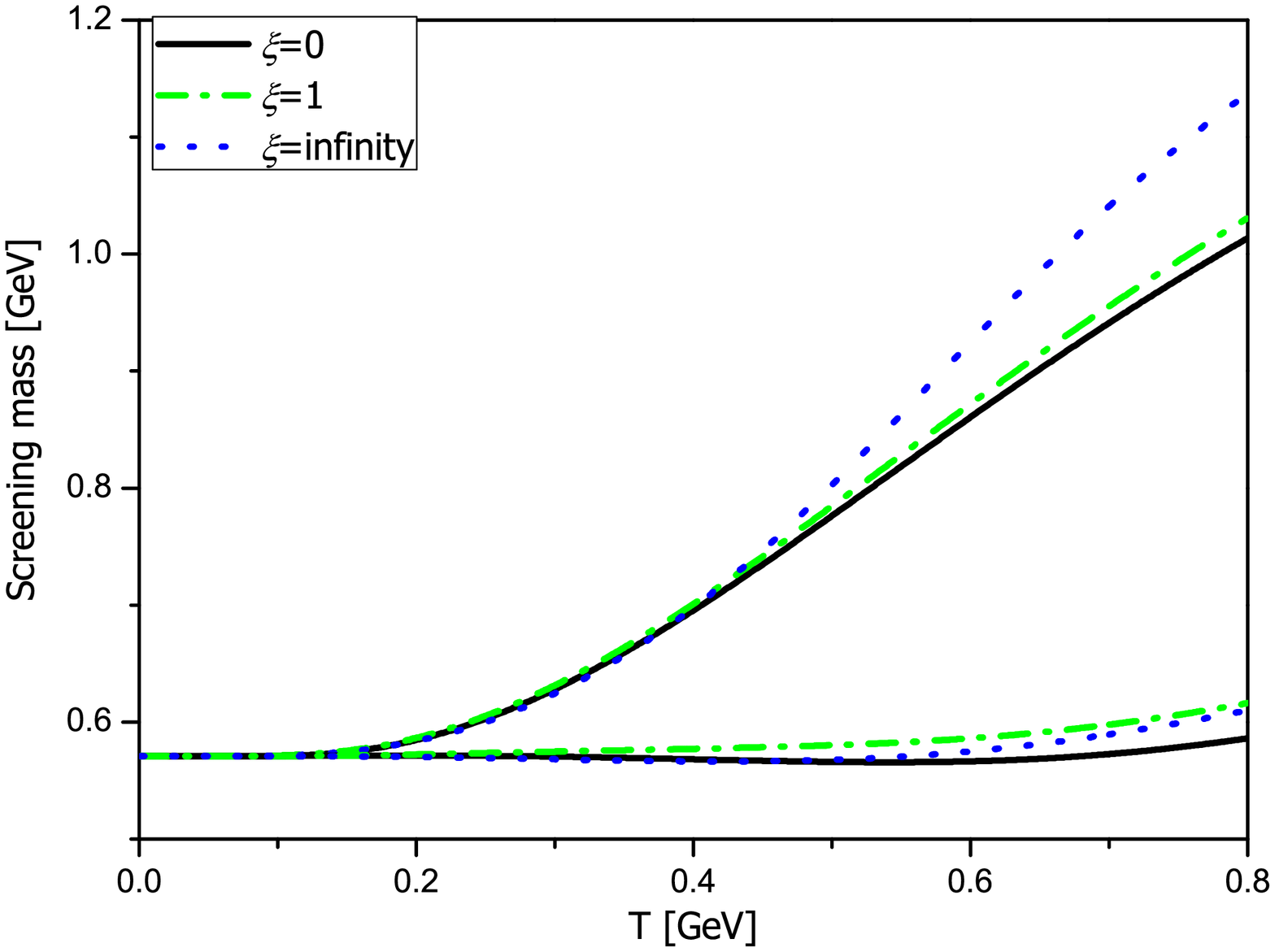}\\
  \caption{The gauge dependence of the screening mass as
  functions of the temperature. }
\label{fig:gauge}\end{figure}

We have shown the screening masses by using the gluon propagator
Eq.(\ref{eq:FT-propg-LG}) in the unitary gauge, i.e. $\xi
\rightarrow \infty$. By fixing the model parameters $\lag
g^2G^2\rag=0.009\times4\pi^2 {\text{GeV}}^4$ and $\Lambda=0.3
[\text{GeV}^{-1}]$, in Fig. \ref{fig:gauge} we show the screening
masses in different gauges, the solid lines are for the the Landau
gauge $\xi=0$, the dash-dotted lines are for the Feynman gauge
$\xi=1$, and the dotted lines are for the unitary gauge $\xi
\rightarrow \infty$. It is found that below $T=500 {\rm MeV}$, the
screening masses are independent on different gauges. The gauge
dependence starts to show up when $T>500 {\rm MeV}$, the electric
screening mass is more sensitive to the gauge fixing than the
magnetic screening mass. In the temperature region we are interested
in, both electric and magnetic screening masses are not sensitive to
the gauge fixing.

\subsection{The Polyakov loop expectation value}

The deconfinement phase transition is characterized by the
Polyakov-loop expectation value. The Polyakov-loop is defined as
\begin{equation}
L(x)=\mr{P} {\rm exp} [ig \int_0^{\beta} d\tau A_4({\bf x},\tau)].
\end{equation}
In order to investigate the relation between the dimension-2 gluon
condensate and the deconfinement phase transition, it is necessary
to calculate the Polyakov-loop expectation value. By using
perturbative expansion \cite{Gava:1981qd}, it has been observed in
Ref. \cite{Megias:2005ve} that the Polyakov loop expectation value
is associated with the electric dimension-2 gluon condensate by the
following relation:
\begin{equation}
<L>={\rm exp}[-\frac{g^2<A_{4}^2>}{4N_cT^2}].
\end{equation}
In our model, the electric dimension-2 gluon condensate has a simple
relation with the electric screening mass square, i.e. $<A_{4}^2> =
m_E^2$.

\begin{figure}[h]
\includegraphics[width=10cm]{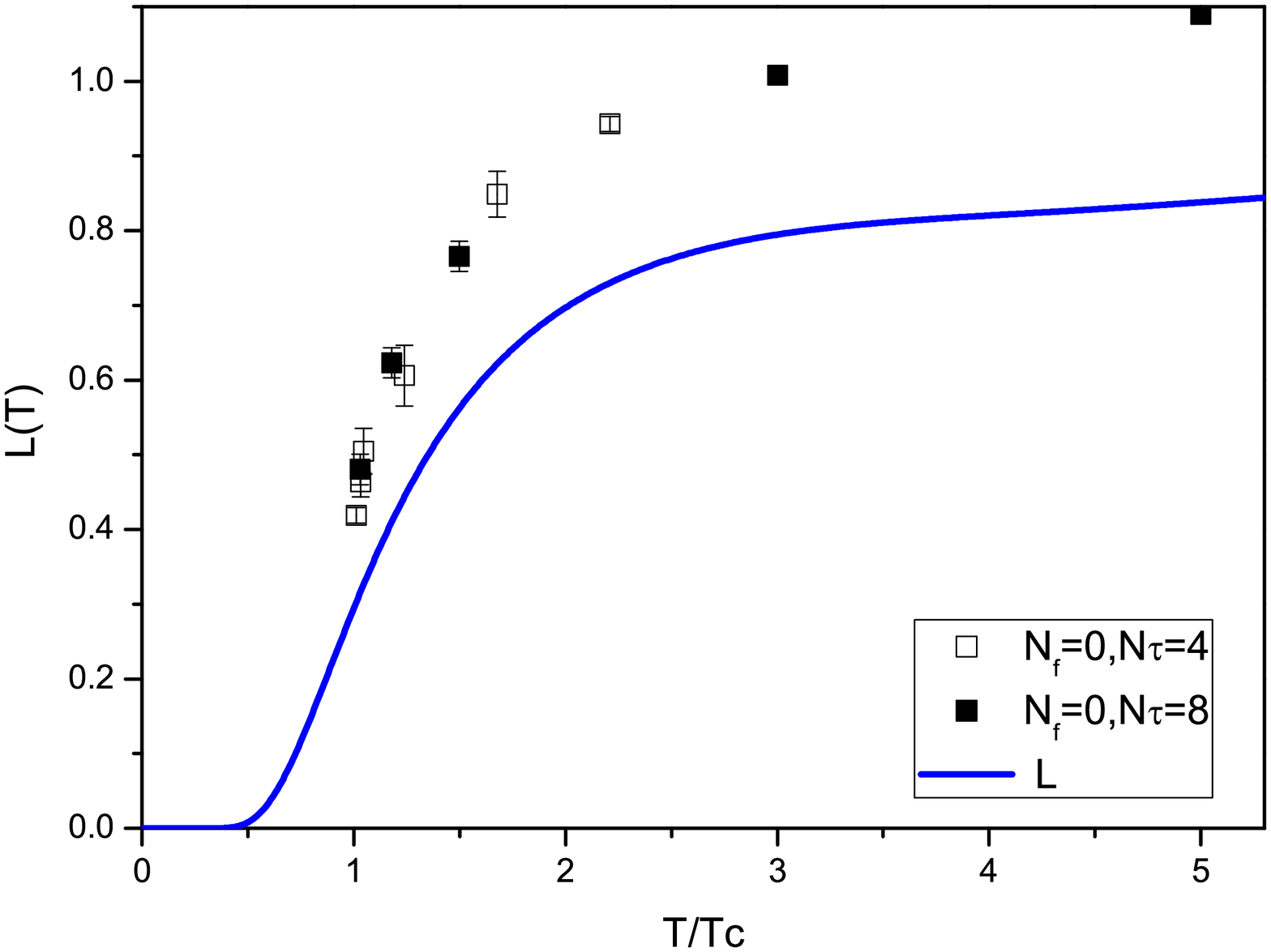}\\
\caption{The Polyakov loop expectation value as a function of
$T/T_c$ comparing with lattice result in
Ref.\cite{Kaczmarek:2002mc}.} \label{fig:L-T}\end{figure}

We show the Polyakov loop expectation value as a function of $T/T_c$
in Fig. \ref{fig:L-T}, and compare the results with lattice data in
Ref.\cite{Kaczmarek:2002mc}. It is found that the Polyakov loop
expectation value is zero in the vacuum and low temperature region,
it starts to rise at around $0.5 T_c$, then rise sharply to a value
of $0.8$ at high temperature. We have taken $T_c=T_0=150 {\rm MeV}$,
where the electric and magnetic gluons start to split. It is worthy
of mentioning that the susceptibility of the Polyakov loop
expectation value indeed gives the critical temperature at around
$T_c=T_0$. Our simple model indicates that the color electric
deconfinement phase transition is driven by the electric gluons, and
the nonperturbative dimension-2 gluon condensate plays an important
role, it still gives at least $80\%$ contribution to the Polyakov
loop expectation value even at temperature region $T>3 T_c$.

\section{Conclusions}\label{sec:concl}

We have investigated the electric and magnetic screenings of the
thermal gluons in a gluodynamic model with dimension-2 gluon
condensate in zero momentum, which spontaneously generates the
effect dynamical gluon mass in the vacuum.

It is found that the electric and magnetic gluons are degenerate at
low temperature. With the increasing of temperature, the electric
and magnetic gluons start to split at certain temperature around
$T_0=150 {\rm MeV}$. The electric screening mass changes rapidly
with temperature at $T>T_0$, and the Polyakov loop expectation value
rises sharply around $T_0$ from zero in the vacuum to a value around
$0.8$ at high temperature. This suggests that the color electric
deconfinement phase transition is driven by electric gluons. It is
also observed that the magnetic screening mass keeps almost the same
as its vacuum value, which manifests that the magnetic gluons
remains confined. Both the screening masses and the Polyakov loop
results are qualitatively in agreement with the Lattice
calculations.

The Polyakov loop expectation value in this work is calculated by
using the perturbation expansion, a more convenient way to derive
the Polyakov loop expectation value is by using AdS/CFT method in
the 5D holographic model, e.g. in Ref. \cite{GCD2-AdS} with a
dimension-2 dilaton field background. It is worthy of mentioning
that the dimension-2 dilaton field corresponds to a dimension-2
gluon condensate operator, and in Ref. \cite{GCD2-AdS}, the Polyakov
loop expectation value at finite temperature agrees well with the
lattice data \cite{Kaczmarek:2002mc}.

The model we used in this paper is quite simple, but it captures
some important feature of gluon dynamics in the vacuum as well as in
at finite temperature. We can conclude that the dimension-2 gluon
condensate plays an essential role both in confinement as well as in
deconfinement phase transition. 

\vskip 1cm \noindent

{\bf Acknowledgments}: The authors thank valuable discussions with
H. Chen, T. Hatsuda, T. Mukherjee, N. Su, Q.S. Yan.  The work of
M.H. is supported by CAS program "Outstanding young scientists
abroad brought-in", CAS key project KJCX2-EW-N01, NSFC under the
number of 10735040 and 10875134, and K.C.Wong Education Foundation,
Hong Kong.

\end{document}